\documentclass[letterpaper]{article}
\usepackage{aaai20}
\usepackage{times}
\usepackage{helvet}
\usepackage{courier}
\usepackage[hyphens]{url} 
\usepackage{graphicx}
\usepackage{graphicx}
\usepackage{amsmath}
\usepackage{amsfonts}
\usepackage{booktabs}

\title{Fast Modeling and Understanding Fluid Dynamics Systems with Encoder-Decoder Networks}

\author{Rohan Thavarajah\textsuperscript{1}, Xiang Zhai\textsuperscript{1}, Zheren Ma\textsuperscript{1} and David Castineira\textsuperscript{1}*  \\ \\ \textsuperscript{1}QRI AI, LLC \\  *david.castineira@qrigroup.com}

\begin{document}
\maketitle
\begin{abstract}
Is a deep learning model capable of understanding systems governed by certain first principle laws by only observing the system's output? Can deep learning learn the underlying physics and honor the physics when making predictions? The answers are both positive. In an effort to simulate two-dimensional subsurface fluid dynamics in porous media, we found that an accurate deep-learning-based proxy model can be taught efficiently by a computationally expensive finite-volume-based simulator. We pose the problem as an image-to-image regression, running the simulator with different input parameters to furnish a synthetic training dataset upon which we fit the deep learning models. Since the data is spatiotemporal, we compare the performance of two alternative treatments of time; a convolutional LSTM versus an autoencoder network that treats time as a direct input. Adversarial methods are adopted to address the sharp spatial gradient in the fluid dynamic problems. Compared to traditional simulation, the proposed deep learning approach enables much faster forward computation, which allows us to explore more scenarios with a much larger parameter space given the same time. It is shown that the improved forward computation efficiency is particularly valuable in solving inversion problems, where the physics model has unknown parameters to be determined by history matching. By computing the pixel-level attention of the trained model, we quantify the sensitivity of the deep learning model to key physical parameters and hence demonstrate that the inversion problems can be solved with great acceleration. We assess the efficacy of the machine learning surrogate in terms of its training speed and accuracy. The network can be trained within minutes using limited training data and achieve accuracy that scales desirably with the amount of training data supplied.
\end{abstract}

\section{Introduction}

In the recent decade, deep learning has demonstrated its power in many different cognitive tasks that were historically believed challenging to conceptualize and quantify with mathematical models, such as image recognition, object detection, semantic segmentation, machine translation, and art generation. With the help of massive datasets \cite{imagenet_cvpr09}, specially designed network architectures \cite{He_resnet} have been very efficient in learning how to perform complex nonlinear feature engineering, capture and master the intrinsic stochastic nature of those tasks, and approximate the human cognitive learning process.

Despite the abundance of success stories in the world of cognitive tasks, the other world of tasks filled with problems and systems that are driven by mathematical logics (i.e., first principle laws, well-defined governing differential equations/models, deterministic or stochastic), still relies on conventional analytical and numerical simulation including finite-difference, finite-volume and finite-element methods. These methods discretize space and time domains into small cells and intervals, transform the partial differential equations into linear and non-linear algebraic problems, and solve those problems numerically. Consider for instance the discrete static 2D Poisson equation which uses simple 3x3 2D kernels $\bigl( \begin{smallmatrix}4 & -1 & 0 \\ -1 & 4 & -1 \\ 0 & -1 & 4\end{smallmatrix}\bigr)$ and $\bigl( \begin{smallmatrix}1 & 0 & 0 \\ 0 & 1 & 0 \\ 0 & 0 & 1\end{smallmatrix}\bigr)$ to convert the second-order partial differential equation into coupled linear systems that can be solved at O($n^3$) time complexity (where $n\times n$ is the size of the discretized 2D grid). In a dynamic system, the time dimension also needs to be discretized and solved sequentially. In computational fluid dynamics (CFD), the time step must be very small to satisfy the CFL condition \cite{CFL_paper}, making CFD time consuming and numerical errors accumulate with time.

In this paper, we show a novel approach of using a physics-based finite-volume simulator to teach domain-lacking deep learning models to accurately simulate 2D subsurface two-phase fluid dynamics in heterogeneous porous media. The governing equations are Navier-Stokes equations plus Darcy's Law 
\begin{equation}\label{eqn:reservoir_pde}
    \frac{\partial}{\partial t}(\phi \rho_i S_i) - \nabla\cdot(\rho_i\lambda_i\mathbf{K}\cdot\nabla p_i) = q_i^w,\ \ i=o, w
\end{equation}
where the subscript $i \in \{o,w\}$ identifies phase as being oil or water, $\rho_i$ is the phase density, $\lambda_i$ is the relative phase mobility, $\mathbf{K}$ and $\phi$ are the permeability and porosity distribution of the porous medium, and $q_i^w$ are the sink/source rates at the locations of producing/injecting wells. The phase saturation $S_i$ (i.e., volumetric percentage of the phase) and pressure $p_i$ are the physics quantities we are trying to solve. Since the two phases are directly connected, we will by default choose the water phase pressure and saturation for discussion. 

When all parameters are given, the dynamics are determined and can be forward simulated using a finite-volume method. In real-world problems, however, the exact form of the equation is not yet determined due to the uncertainty of subsurface rock properties like permeability $\mathbf{K}$. Meanwhile, the system can be measured/observed at some locations (such as wells). A more important and challenging problem is to, in reverse fashion, estimate the physics parameters $\mathbf{K}$ based on the observed data. In CFD approaches, a forward simulation is conducted with an initial guess of the permeability map and simulated behavior at observable locations is compared with real measurements. The initial guess of permeability is then revised in order to yield new behavior that better aligns with real data. This iterative revision process is called history-matching and is an example of an inverse problem wherein one wishes to infer the exact form of the governing laws based on limited observed data. Forward evaluation is itself expensive, inverse problems, which require many forward evaluations, are much more demanding. We will show that the deep-learning-based surrogate model and its interpretation offer unprecedented advantages to solve the inverse problem.

There have been several recent efforts to capture the physics of a system using fast deep learning surrogates. Examples appear in rainfall prediction, animation, aerodynamics design and of particular interest to us, reservoir simulation \cite{shi_convolutional_nodate,ladicky_data-driven_2015,guo_convolutional_2016,zhu_bayesian_2018}. These efforts may be summarized according to two broad strategies.

\begin{enumerate}
\item Data-Driven Approaches; the network observes the system and derives the physics from data. Our approach is data-driven in nature.
\item Physics-Embedded Approaches; a scientist supplies the governing physics equations directly to the network and the network is trained to adhere to them.
\end{enumerate}

Inspired by the success of CNN architectures on image translation problems like pix2pix \cite{isola_pix2pix_2016}, \citeauthor{zhu_bayesian_2018} \shortcite{zhu_bayesian_2018} introduce a fully convolutional encoder-decoder network (DenseED) to model the behavior of fluid in heterogeneous media. Encoder-decoder networks are often motivated by an intuition that input and output images share an underlying structure. However, though no such structure is readily apparent, \citeauthor{zhu_bayesian_2018} \shortcite{zhu_bayesian_2018} find that DenseED successfully translates permeability to steady-state velocity fields. \citeauthor{mo_integration_2019} \shortcite{mo_integration_2019} extend DenseED to incorporate GAN loss to tackle highly non-linear outputs (for other examples of GANs applied to proxy particle shower simulations and heat conduction see \citeauthor{paganini_accelerating_2018} \shortcite{paganini_accelerating_2018} and \citeauthor{farimani_deep_2017} \shortcite{farimani_deep_2017} respectively). 

Time dependence is captured using a variety of strategies. Without altering the architecture of DenseED, \citeauthor{mo_deep_2019} \shortcite{mo_deep_2019} model time-dependence by broadcasting time across an additional input channel. By treating time in this manner, \citeauthor{mo_deep_2019} are able to furnish predictions at arbitrary time instances. However, fluid flow is highly autoregressive in nature; the state of the system at time $t$ is a function of the system's state at $t-1$ and \citeauthor{mo_deep_2019} do not explicitly exploit this structure available in the data. In contrast \citeauthor{wiewel_latent-space_2018} \shortcite{wiewel_latent-space_2018} combine an encoder-decoder with a long short-term memory network (LSTM) and task the LSTM with learning transitions in the encoder-derived latent space. \citeauthor{shi_convolutional_nodate} (2015) and \citeauthor{wang_predrnn:_nodate} (2017) further customize LSTMs for spatiotemporal data by innovating layers that respectively include convolution operations in transition functions and construct a shared memory pool across LSTM layers.

Alternatively, physics-embedded approaches take an unsupervised approach and utilize the governing equations of a system directly. \citeauthor{raissi_physics_2017} \shortcite{raissi_physics_2017} prescribe a framework for embedding physics into a neural network by expressing the loss as the sum of two terms, one that describes the dynamics of the system and another that describes boundary conditions. As \citeauthor{raissi_physics_2017} minimize this loss, they find a function approximation that satisfies known dynamics and boundary conditions and so solve the PDE directly.  \citeauthor{zhu_physics-constrained_2019} \shortcite{zhu_physics-constrained_2019} demonstrate how spatial gradients can be computed using Sobel filters and therefore extend the framework pioneered in \citeauthor{raissi_physics_2017} to convolutional architectures.


The paper is organized as follows. First, we formulate the problem and describe deep neural network architectures that have successfully learned from a CFD simulator. We compare two different approaches to dealing with time dependence and discuss how accuracy and computation time of the deep learning surrogate scale with more training data. Then we show how training with adversarial loss is helpful for solutions with large spatial gradient. Finally, we demonstrate how the model explanation can convert the inverse problem into a gradient-enabled optimization problem that can be solved 4-5 orders of magnitude faster than traditional numerical methods.

\begin{figure*}[t!]
\centering
\includegraphics[width=5in]{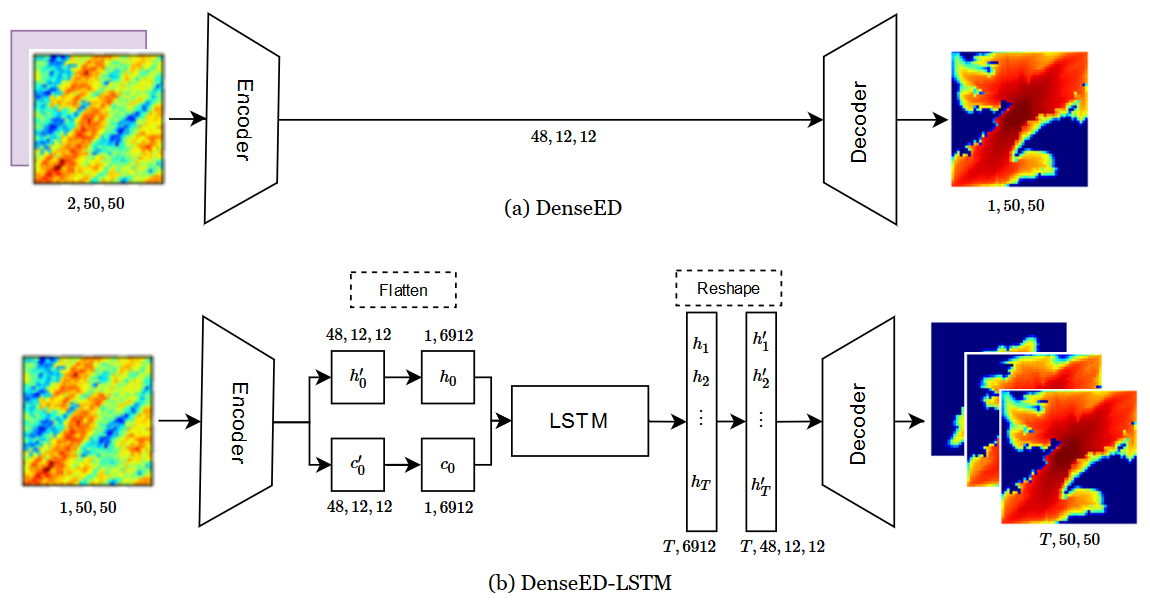}
\caption{DenseED models time by broadcasting it across a 50x50 input channel (purple) so that it receives two-channel input. DenseED-LSTM receives a single permeability channel as input and outputs $T$ saturation fields. We train both models using Adam as our choice of optimizer, MSE loss and an initial learning rate of $3e^{-3}$ which we reduce using a scheduler over 200 epochs.}
\label{fig:architecture_compressed}
\end{figure*} 

\section{Method}

\subsection{Problem Formulation}
We wish to develop a fast, accurate surrogate to a CFD simulator that solves a collection of PDEs (Eqn \ref{eqn:reservoir_pde}) given a permeability distribution $\mathbf{K}$ on a uniform 2D Cartesian grid. Since $\mathbf{K}$ is the input of the system, we represent $\mathbf{K}$ by 2D tensor $X \in \mathbb{R}^{1\times H\times W}$ where $H$ and $W$ denote the height and width of the grid. In general, we can extend $X$ to 3D and include multiple inputs by expanding the tensor $X \in \mathbb{R}^{n_x\times H\times W\times D}$ where $n_x$ counts the number of input types and $D$ denotes depth.

The output of the simulator, $Y$, is spatiotemporal and describes the path of two maps; a saturation map that describes the relative composition of water in each cell, and a pressure map. Again, we represent $Y \in \mathbb{R}^{T\times n_y\times H\times W}$ as a tensor, where $n_y$ counts the total number of output maps ($n_y=2$) and $T$ is the total number of time steps. The simulator can be represented as a function that maps $f:X \rightarrow Y$. We cast the problem as an image-to-image regression treating $n_x$ and $n_y$ as channels and $H\times W$ as the shape of the input and output images.

We nondimensionalize Eqn \ref{eqn:reservoir_pde} and choose $H=W=50$. Initially, the entire 50 by 50 region is filled with oil, and hence water saturation starts as zero everywhere. An injector placed at the center of the region injects water at a constant rate (source term in Eqn \ref{eqn:reservoir_pde}), replacing and pushing the liquid towards four constant-rate producers at the four corners. This problem setting describes a waterflood, a practice of secondary recovery of crude oil extraction, in which water, less viscous and immiscible with crude oil, is injected into a reservoir to achieve higher long-term ultimate oil recovery as well as maintain subsurface pressure.

We create a prior permeability map using a sequential Gaussian method and divide the region into six sub-regions based on its value and pixel proximity. We run 400 CFD simulations to time step $T=30$ wherein each simulation is characterized by a different permeability map $X$ generated by multiplying six randomly sampled independent multipliers in the range $[e^{-2},e^{2}]$ to the six sub-regions of the prior permeability map. We effectively obtain a 400-case dataset filled with synthetic data that correspond to different physics input. $336$ cases are used as the training set, and the remaining $64$ cases are used for testing. The goal is to train networks with the training simulation data that yield reliable predictions of pressure and saturation on the hitherto unseen permeabilities in the test set. 

As we later discuss, we recognize that pressure is a more global property than saturation, demanding a larger receptive field. Furthermore, relative to saturation pressure exhibits less variation over time. Owing to these diverging properties, we decouple the problem and separately predict saturation and pressure.

\subsection{Architecture}





We employ the network architecture developed by \citeauthor{zhu_bayesian_2018} \shortcite{zhu_bayesian_2018} (DenseED) and incorporate an LSTM as an alternative treatment of time and GAN loss to better simulate shocks. The design presented in \citeauthor{zhu_bayesian_2018} \shortcite{zhu_bayesian_2018} is a combination of two structures; \textbf{convolutional encoder-decoders} and \textbf{Densenet}. Encoder-decoder networks have successfully been applied to image-image translation problems such as image segmentation and pattern infilling \cite{long_fully_nodate,badrinarayanan_segnet_2015}. Encoder-decoders subject an image to a coarse-refine process. With each successive encoding layer, the network coarsens, extracting higher-level features from the image at lower spatial resolution. Ultimately, the encoder finds an underlying representation of the image in a latent space. With each successive decoding layer, this underlying representation is refined to construct the output. 

\begin{figure}
\centering
\includegraphics[width=2.4in]{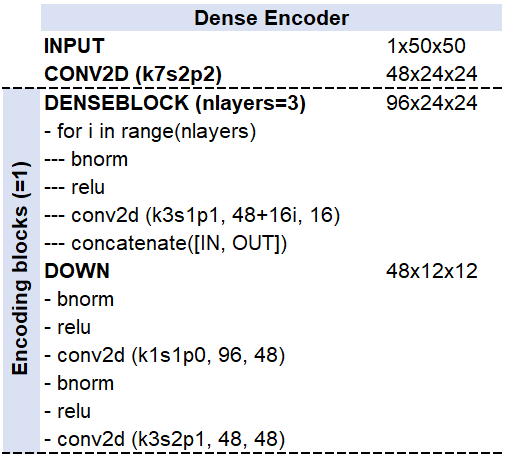}
\caption{Encoder Design. We represent the arguments of the convolution operations in parentheses. Conv2d (k3s1p1, 48, 16) denotes a 2D convolution with kernel size 3, stride 1, padding 1, which receives 48 features maps as input and outputs 16. DenseED is modularized using encoding and decoding block units which consist of a dense block and a transition block. Each encoding (decoding) block halves (doubles) the dimensions of the output feature map. We find that encoding down to 12x12 features maps works best for saturation and encoding down to 6x6 works best for pressure.}
\label{fig:encoder_v2}
\end{figure}

\begin{figure}
\centering
\includegraphics[width=2.4in]{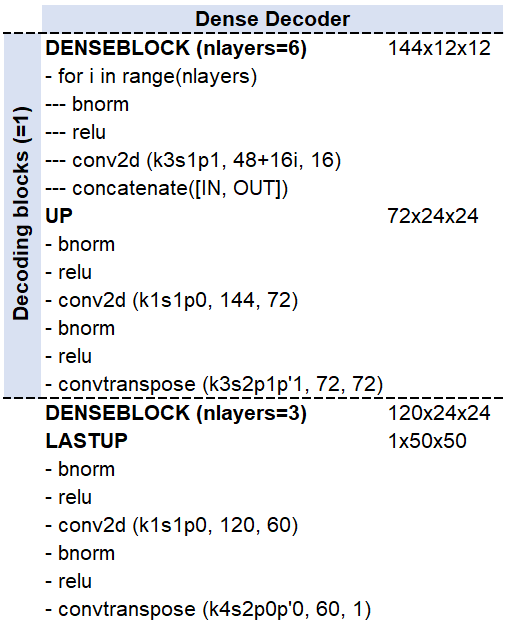}
\caption{Decoder design. See encoder for details.}
\label{fig:decoder_v2}
\end{figure}

DenseNets \cite{huang_densely_2016} choose convolutions that preserve identical dimensions between inputs and outputs, thereby permitting their concatenation. The input to any one layer is the last layer's output concatenated with all previous inputs. This structure is particularly valuable when analyzing objects across multiple spatial scales. Each successive convolution reflects an expanding receptive field. A 1x1 convolution combines feature maps of varying receptive fields resulting in a network that readily adapts to local versus global phenomena. We present the structure of DenseED in Panel A of Figure \ref{fig:architecture_compressed} and detail the specific design of the encoder and decoder in Figures \ref{fig:encoder_v2} and \ref{fig:decoder_v2}.

\citeauthor{mo_deep_2019} \shortcite{mo_deep_2019} propose a strategy for modeling time-dependent outputs without altering the design of DenseED by broadcasting time across an input channel. We experiment with incorporating an LSTM \cite{hochreiter_long_1997} to model transitions in the encoder-derived latent space henceforth referred to as DenseED-LSTM (Panel A of Figure \ref{fig:architecture_compressed}). We hypothesize that an LSTM, with its memory and transition properties, is particularly apt to model a simulator's dynamic behavior. The persistent cell state will house information about the permeability grid, which is relevant to prediction at every time step. Meanwhile, the hidden state-to-state transitions will receive a strong signal due to the autoregressive nature of the governing physics. Compared to DenseED, the encoder in DenseED-LSTM outputs twice as many feature maps (96 vs. 48). Half of those feature maps are used to initialize the first hidden state and half the first cell state. The LSTM outputs hidden state vectors for $T$ time steps which we decode to predict the saturation path.

\section{Saturation Solutions and Alternative Treatment of Time-Dependence}

\begin{table}
\centering
  \caption{Goodness of fit}
  \label{table:gof}
\resizebox{0.75\columnwidth}{!}{
  \begin{tabular}{lcc}
    \toprule
     & Train $R^2$ & Test $R^2$ \\
    \midrule
     DenseED & 0.991 & 0.840 \\
     DenseED-LSTM & 0.941 & 0.847 \\
  \bottomrule
\end{tabular}}
\end{table}

We evaluate the performance of the surrogate in predicting the path of saturation with $R^2 = 1-\sum_{i=1}^{N} (\hat{y_i}-y_i)^2 / (y_0-\bar{y})^2$, where $y_i$ is a tensor describing the true saturation path for simulation $i$ and $\hat{y_i}$ is the corresponding model prediction. The null model, from which we derive $\bar{y}$, predicts the mean image for all time steps. In Table \ref{table:gof}, we record the test performance of DenseED and DenseED-LSTM which achieve similar degrees of success.

\begin{figure}
\centering
\includegraphics[width=2.6in]{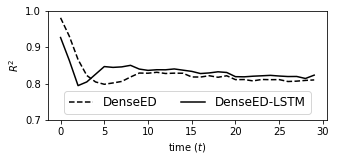}
\caption{Test $R^2$ evaluated at each time step. Both DenseED and DenseED-LSTM exhibit initially strong performance that deteriorates to a floor.}
\label{fig:r2}
\end{figure}

We further characterize the model's performance by exploring how error varies in time and space. We expect time will influence error because the difficulty of the prediction task is non-uniform. Initially easy, prediction becomes difficult and then reverts to being easy again over short-, mid- and long-time horizons. The first images are easy to predict because water is initially only concentrated around the injector. As water floods the field, saturation enters a period of rapid flux. In this period, saturation at $t-1$ bears the least resemblance with its state at $t$, and prediction is commensurately more difficult. Once saturation steadies, the model can issue a prediction at time $t$ at least as accurately as it did at $t-1$. Therefore we expect performance to plateau at long time horizons.  We observe that for both DenseED and DenseED-LSTM, $R^2$ follows this trend (Figure \ref{fig:r2}). Performance is highest when $t$ is near zero and deteriorates to a floor.

\begin{figure}
\centering
\includegraphics[width=3in]{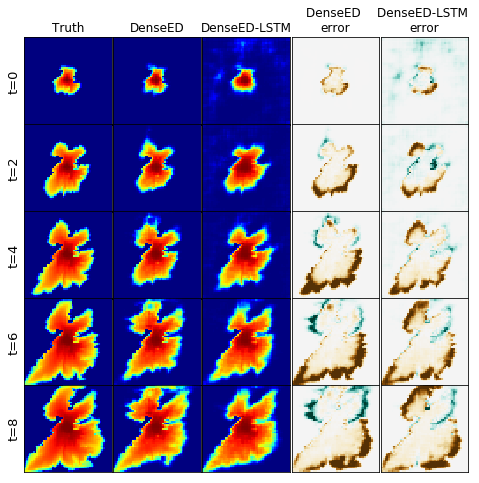}
\caption{Path of saturation and predictions for a single simulation in the test set. A conventional solver provides ground truth. DenseED and DenseED-LSTM are trained using 334 simulations. Saturation is a proportion that takes values between 0 (blue) and 1 (red). We plot error using a diverging color palette and enforce symmetry by setting vmin, vmax = -0.5 (brown), 0.5 (blue).}
\label{fig:sat_path}
\end{figure}

We also inspect individual test predictions to examine the spatial distribution of error. In Figure \ref{fig:sat_path}, we present one such prediction. We observe that for both DenseED and DenseED-LSTM error tends to be concentrated around the saturation front. 

\begin{figure}
\centering
\includegraphics[width=3in]{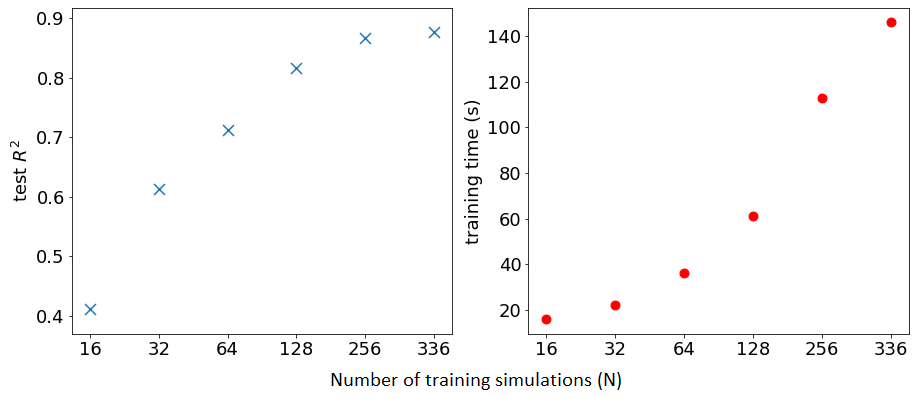}
\caption{Scaling. DenseED achieves high performance with limited training data.}
\label{fig:scaling}
\end{figure}

Once trained, the neural network can make very fast forward predictions. The CFD simulator took 10 seconds to finish each simulation when creating the synthetic dataset on an Intel Xeon 2.4GHz 128GB RAM Nvidia K80 workstation. For comparison, the neural network makes batch predictions on 1000 different permeability maps in less than a second on the same machine. The main time overhead of the neural network is in training. Still, it only takes 2 minutes to train DenseED for 200 epochs to predict a specific time instance with a Nvidia K80. In Figure \ref{fig:scaling}, we show how test performance and training time vary with the number of simulations provided in training. Unsurprisingly the more simulations are provided in training, the better the model's performance on the test set.

\section{Global properties and Adversarial Effect}
Unlike saturation which evolves at the speed of mass motion, pressure is a "global" physics property established at the speed of sound, several orders of magnitude faster than the fluid's convection speed. Therefore, the pressure solution is affected by the permeability of the entire region. In other words, the receptive field of pressure is global. Indeed, we find that we obtain the best validation results by adding one additional encoding and one additional upsampling block to the DenseED structure, which shrinks the bottleneck of the autoencoder from $12\times12$ to $6\times6$. The smaller bottleneck further contains six $3\times3$-kernel convolutional layers, effectively resulting in a global receptive field of each pixel at the end of the bottleneck.

\begin{figure*}
\centering
\includegraphics[width=5.5in]{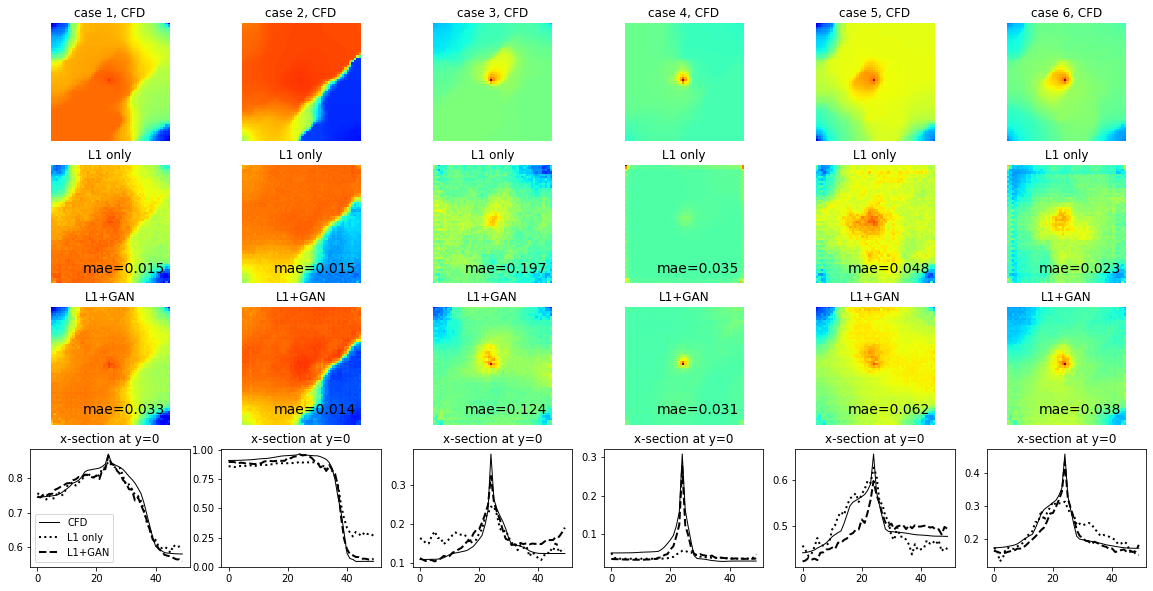}
\caption{Pressure solutions scaled to $[0,1]$ at time $t=25$ of six randomly sampled permeability maps from the test dataset. Top row: True solution from numerical CFD. 2nd row: Outputs from DenseED trained with L1 loss only. 3rd row: Outputs from the same DenseED architecture trained with L1 and GAN loss. Both the L1-only and L1+GAN cases are trained using an initial learning rate of $5\times10^{-4}$ that is halved every $30$ epochs over $250$ total epochs. The relative weight $\alpha$ for the GAN loss term is $50$. The discriminator is trained using soft labels ($0.9$ and $0.1$). The median absolute errors are listed for each of the L1-only and L1+GAN solutions.  Bottom row: Cross-sectional view of the three solutions at $y=0$.}
\label{fig:pressure_gan}
\end{figure*}

Fluid dynamic systems are known to have sharp spatial gradient and discontinuity like shocks. For example, the pressure solution scales as $\log(r)$ near the sources and sinks in 2D or cylindrical 3D space. This can lead to enormous spatial gradients close to the wells. Vanilla generative neural networks have difficulty in capturing such behavior because the regular distance-based losses lead to smoother, blurry results in an attempt to resemble many possible solutions. In the case of a fluid dynamic system, although the solution is unique for a given permeability map, we still found large-gradient results are more difficult to generate (see the 2nd row of Figure \ref{fig:pressure_gan}).

Generative Adversarial Networks (GAN) \cite{goodfellow_gan} are well-known for creating sharp and visually realistic results by introducing a smart loss term learned by comparing the real and generated results. An additional discriminator $D$ is trained simultaneously to penalize the generator if it produces results that the discriminator easily discerns as fake. Similarly to \citeauthor{isola_pix2pix_2016} \shortcite{isola_pix2pix_2016},  we adopt the final objective as
\begin{equation}\label{eqn:gan_goal}
    G^* = \mathcal{L}(G) + \alpha \cdot \arg \min_G\max_D \mathcal{L}_{GAN}(G,D),
\end{equation}
where $G$ is the DenseED autoencoder (generator), $D$ is the discriminator which has the same architecture as the encoder of DenseED plus a fully-connected layer to output binary prediction, and $\mathcal{L}$ is the distance-based loss (L1-loss by default) between pixels of generated results and CFD results (true label). The adversarial loss term is given by
\begin{equation}\label{eqn:gan_loss}
    \mathcal{L}_{GAN}(G,D) = \mathbb{E}_{y}[\log D(y)] + \mathbb{E}_x [\log(1-D(G(x))],
\end{equation}
where $x$ is the permeability map input to DenseED, $y$ is the CFD result, and $G(x)$ is the generated result. In \citeauthor{isola_pix2pix_2016} \shortcite{isola_pix2pix_2016}, a conditional discriminator is utilized to judge if an image is real and also relevant to the input $x$. In our case, although we do not find the conditional adversarial mechanism enhances the results notably, it does help stabilize the GAN training process. A fundamental difference between Eqn \ref{eqn:gan_loss} and regular GAN is the absence of a randomly sampled latent vector. Once the permeability map $x$ is given, the fluid system's dynamics are determined. Therefore, unlike a traditional GAN, we do not feed a random latent vector $z$ as input to the generator. The term $\alpha$ in Eqn \ref{eqn:gan_goal} is a weight quantifying the relative importance of GAN loss and is chosen to be $50$ so that the GAN loss is about 50\% of L1 loss at the conclusion of the training cycle.

Figure \ref{fig:pressure_gan} compares the pressure solutions from CFD, DenseED trained with L1 loss only and the same DenseED architecture trained with both L1 and GAN loss (Equation\ref{eqn:gan_loss}). GAN loss helps significantly in capturing the large pressure gradient near the central injector. In comparison, L1-only DenseED produces consistently smoother results. Although the overall pixel-wise error of L1-only DenseED is smaller in many cases, the L1+GAN results are scientifically more accurate near the sink/source locations. This is critical because, in the real world, the sink/source locations are where the observable data are sampled.

\section{Model Explanation and the Inverse Problem}
Interpreting complex machine learning models such as deep neural networks is important because it helps humans understand the behavior of the model, vet whether the prediction is made based on the correct reasoning, and build even better models. Pixel-wise sensitivity and/or spatial attention maps have been used to help understand convolutional neural networks \cite{Bach2015,smoothgrad_cnn,SHAP,Zagoruyko2016_attention}. Given that the deep learning models have been proven capable of accurately mimicking CFD numerical simulations, one important application is to explain the neural network model and uncover its underlying logic.

Considering saturation $S$ as a function of permeability $\mathbf{K} = K(x',y')$ and time $t$, the "pixel-wise" sensitivity of saturation to the permeability distribution can be quantified by
\begin{equation}\label{eqn:sensitivity_def}
    X(x,y,t,x',y',\mathbf{K}) \equiv \frac{\partial S(x,y,t,\mathbf{K})}{\partial K(x',y')}.
\end{equation}
The partial in Equation\ref{eqn:sensitivity_def} represents how much saturation at location and time $(x,y,t)$ would change if the permeability at $(x',y')$ were a bit higher. In real cases, $(x,y,t)$ are the locations/time where the system is observed/measured. Given a fixed location of interest $(x_0, y_0)$ and a time $t_0$, the local saturation sensitivity to permeability is a 2D slice of Eqn \ref{eqn:sensitivity_def}: 
\begin{equation}\label{eqn:sesitivity_local}
  \mathbf{X}_{x_0,y_0,t_0,\mathbf{K}} \equiv X(x',y')_{x_0,y_0,t_0,\mathbf{K}} =  \frac{\partial S(x_0,y_0,t_0,K)}{\partial K(x',y')}.
\end{equation}

The sensitivity map helps us understand what consequences a slight alteration of the input physics laws may have on the system. However, the sensitivity map $\mathbf{X}$ is extremely difficult to estimate using traditional CFD approaches because there is no explicit formula of $S(x,y,t,\mathbf{K})$. Instead, sensitivity must be calculated by forward computation at very high cost. In the given case, in order to numerically estimate $\mathbf{X}$ on a 50x50 grid at time $t=25$, $2500$ simulations have to be run till this time step for each slightly perturbed input permeability map, which would take 7 hours on the same workstation. When the permeability distribution $\mathbf{K}=K(x',y')$ has to be updated iteratively, say, 100 times, in the inversion problem, the CFD-based approach would take more than one month.

Thanks to the deep-learning surrogate model, $\mathbf{X}$ can be computed directly via backward propagation. Since DenseED and the convLSTM takes $\mathbf{K}$ as input, the derivative of the output of the networks w.r.t. the "pixels" of the input permeability map is the same as Eqn \ref{eqn:sesitivity_local}. We can also compute $\mathbf{X}$ easily via numerical differentiation by running a batch of $2500$ forward prediction at low cost. For the same example, evaluating six 2D $\mathbf{X}$ at six different $(x_i,y_i,t_i)$ on the same workstation took only 619 ms, a 40,000-fold acceleration.

\begin{figure}
\centering
\includegraphics[width=3in]{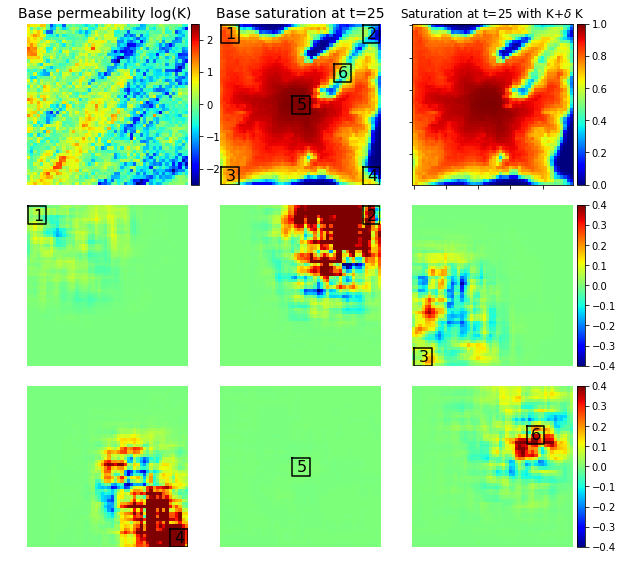}
\caption{Top rows: Input permeability map (left) and the saturation solution given by DenseED at time $t=25$ (mid). Top right shows the saturation solution with a slightly perturbed permeability. Row 2-3: Pixel-wise sensitivity at six observation sites to the input permeability. A value $\beta$ at coordinate $(x,y)$ in the sensitivity map for Location-\textbf{i} means that a $\Delta k$ increase in the permeability map at $(x,y)$ will result in a $\beta\Delta k$ increase in the saturation at location \textbf{i} at time $t=25$. The partial derivatives are computed by directly backpropagating DenseED and averaged via a $5\times 5$ kernel.}\label{fig:sensitivity}
\end{figure}

Figure \ref{fig:sensitivity} shows the saturation sensitivity map at six locations at time $t=25$ for a given permeability map from the test set. The particular chosen case is more permeable on the left half of the region, resulting in earlier water breakthrough in locations 1 and 3 than locations 2 and 4. As a result, the upper right and lower right corners have higher sensitivity to the permeability distribution between them and the central injector. The central point has zero sensitivity to the permeability map because it is always flooded immediately regardless. Location 6, which is located in a low permeability channel, is most sensitive to its upstream, the small region between itself and the central injector. In some sensitivity maps, certain regions have negative impact, indicating that an increase of permeability in those regions will actually result in a lower saturation; this is because those regions with a higher permeability will redirect some water to other directions and consequently give lower saturation at the locations of interest. The sensitivity maps, effectively obtained by locally linearizing the deep-learning-based surrogate model, honor the underlying physics very well, confirming the deep-learning model's ability to capture the physics by learning from data.

The interpretation of the deep learning model lends itself to a powerful application. In the inverse problem, a dynamic system is governed by certain but not-yet-fully-determined laws. The goal is to find out the exact form of the governing laws so that the resulting dynamics satisfy all the observed data. In our case, the saturation at $(x_i,y_i,t_i)$ is observed and certain; however, the permeability distribution, which gives the exact form of the fluid dynamic equations, is not yet determined. The solution should be an optimal $\mathbf{K}^* = K(x',y')$ so that the resulting $S$ has minimized total discrepancy with all the observed data, i.e.,
\begin{equation}
\mathbf{K}^* = \underset{\mathbf{K}}{\mathrm{argmin}} \mathcal{L}_{\text{IP}},\ \
\mathcal{L}_{\text{IP}} = \Sigma_i || S(x_i,y_i,t_i,\mathbf{K}) - S_i||
\end{equation}

where $\{S_i\}$ are the set of real measurements at locations/times $(x_i,y_i, t_i)$. With L2 norm, the loss of the inverse problem is differentiable w.r.t. $\mathbf{K}$:
\begin{equation}
\nabla_{\mathbf{K}}\mathcal{L}_{\text{IP}} = 2 \Sigma_i ( S(x_i,y_i,t_i,\mathbf{K}) - S_i)\cdot \mathbf{X}_{x_i,y_i,t_i,\mathbf{K}}
\end{equation}
Given that $\mathbf{X}$ can be readily evaluated with the deep-learning surrogate model, the above problem can be solved using gradient-based optimization methods iteratively via
$$\mathbf{K}\leftarrow \mathbf{K} - \gamma\cdot\nabla_{\mathbf{K}}\mathcal{L}_{\text{IP}},$$
where $\gamma$ is the learning rate in the above gradient-descent iteration. In other words, in every iteration of the inversion optimization, the input permeability map is added/subtracted by a linear combination of the sensitivity maps shown in Figure \ref{fig:sensitivity}, and then the sensitivity maps are recomputed to adjust for the change of the input permeability distribution. For instance, we perturbed the input permeability by adding $0.1$ times the sensitivity map 2 and subtracting $0.3$ times the sensitivity map 3, to obtain a desired saturation map (upper right corner of Figure \ref{fig:sensitivity}) in which location 2 has higher saturation and location 4 has lower saturation.

We show in this section that model explanation offers an unprecedented advantage in solving the inverse problem, the true challenge that is not yet achievable with the traditional CFD method due to the limitation of computation power.

\section{Summary and Discussion}
We have presented a novel method of modeling a special fluid dynamic system with deep learning. We show that, by sampling a good distribution of exact form of the physics laws, it is possible to teach one unique deep learning model to capture the underlying physics and accurately simulate the dynamic system under each of the sampled physics law, and make prediction at high accuracy to new physics that were not explicitly provided in the training data; moreover, the deep learning approach provides a significantly faster way to forward evaluate the system, potentially helping solve the motivating inverse problem. The nonlinear partial dependence of the system behavior to the imposed physics can be quantified using the techniques of neural network explanation. This provides a much easier way to solve the inverse problem that would otherwise be infeasible using  traditional numerical methods due to their high cost.



\nocite{*}
\bibliography{sample-bibliography}
\bibliographystyle{aaai}

\end{document}